\documentclass[superscriptaddress,
twocolumn,secnumarabic,nofootinbib]{revtex4-1}
\usepackage{amsmath}
\usepackage[utf8]{inputenc}
\usepackage[T1]{fontenc}

\pdfoutput=1

\usepackage{bm}
\usepackage{times}
\usepackage{braket}
\usepackage{color,graphicx}
\usepackage[utf8]{inputenc}
\usepackage[T1]{fontenc}
\usepackage[export]{adjustbox}

\usepackage{enumitem}

\usepackage{verbatim}

\usepackage{float}

\usepackage{colortbl}
\usepackage{longtable}
\usepackage{tabularx}
\usepackage{makecell}

\usepackage{multirow}

\usepackage{xcolor}

\usepackage{lipsum}

\usepackage[normalem]{ulem}

\definecolor{LightGray}{gray}{0.96}
\definecolor{Gray}{gray}{0.94}
\definecolor{nicered}{rgb}{0.7,0.1,0.1}
\definecolor{nicegreen}{rgb}{0.1,0.5,0.1}

\usepackage[colorlinks,
            linkcolor=black,
            filecolor=black,
            anchorcolor=black,
            urlcolor=black,
            citecolor=blue,
            bookmarks=false,
            ]{hyperref}

\allowdisplaybreaks

\setlongtables

\begin{document}

\title{\boldmath The influence of invisible light particles on $\Lambda_b \to \Lambda E_{\mathrm{miss}}$}

\author{Quan-Yi Hu} \email{huquanyi@gxnu.edu.cn}
\author{Zhi-Bin Duan} 
\affiliation{Department of Physics, Guangxi Normal University, Guilin 541004, Guangxi, China}
%\affiliation{Guangxi Key Laboratory of Nuclear Physics and Technology, Guangxi Normal University, Guilin 541004, Guangxi, China}
  
%%%%%%%%%%%%%%%%%%%%%%%%%%%%%%%%%%%%%%%%%%%%%%%%%%
\begin{abstract}
In this work, we study the contribution of invisible light particles to $\Lambda_b \to \Lambda E_{\mathrm{miss}}$, particularly the three-body decays $\Lambda_b \to \Lambda \phi \bar\phi$ and $\Lambda_b \to \Lambda \psi \bar\psi$. The differential branching ratio of $\Lambda_b \to \Lambda E_{\mathrm{miss}}$ and the $q^2$-dependent longitudinal polarization asymmetry of $\Lambda$ in scenarios explaining the Belle~II excess are presented. In the chiral basis, we investigate the correlations between $\mathcal{B}(B \to K E_{\mathrm{miss}})$ and $\mathcal{B}(\Lambda_b \to \Lambda E_{\mathrm{miss}})$, as well as between $\mathcal{B}(B \to K E_{\mathrm{miss}})$ and $P^{\Lambda}_{L}$, in eight distinct new physics scenarios. We find that the $P^{\Lambda}_{L}$ can be used to distinguish the chirality of the hadronic current part in the effective operators, which is similar to the cases in the two-body decays $\Lambda_b \to \Lambda \phi$ and $\Lambda_b \to \Lambda V$.
\end{abstract}
%%%%%%%%%%%%%%%%%%%%%%%%%%%%%%%%%%%%%%%%%%%%%%%%%%
\maketitle

\section{Introduction}
\label{sec:introduction}

Rare $b \to s \nu \bar{\nu}$ transitions are flavour-changing neutral current (FCNC) processes absent at tree level and strongly suppressed at higher orders by the Glashow-Iliopoulos-Maiani (GIM) mechanism~\cite{Glashow:1970gm} in the Standard Model (SM), making them sensitive to new physics (NP). Recently, the Belle~II collaboration has found the first evidence of the $B^+ \to K^+ E_{\mathrm{miss}}$ decay. The branching ratio of this decay was measured using two different techniques: the hadronic-tagged analysis and the inclusive-tagged analysis. The combined result is as follows~\cite{Belle-II:2023esi}
\begin{align}
\label{eq:BKexp}
\mathcal{B}(B^+ \to K^+ E_{\mathrm{miss}})_{\mathrm{exp}} = (23 \pm 7) \times 10^{-6}.
\end{align}
In the SM, the missing energy in the final state is carried by a pair of massless neutrinos, with the corresponding branching ratio being~\cite{Hu:2024mgf}
\begin{align}
\label{eq:BKsm}
\mathcal{B}(B^+ \to K^+ \nu \bar{\nu})_{\mathrm{SM}} = (5.09 \pm 0.41) \times 10^{-6}.
\end{align}

The Belle~II measurement is approximately $2.6 \sigma$ higher than the SM prediction. This discrepancy may be caused by NP beyond the SM~\cite{Athron:2023hmz,Bause:2023mfe,Allwicher:2023xba,Abdughani:2023dlr,Chen:2023wpb,He:2023bnk,Datta:2023iln,Altmannshofer:2023hkn,McKeen:2023uzo,Fridell:2023ssf,Ho:2024cwk,Chen:2024jlj,Gabrielli:2024wys,Hou:2024vyw,Chen:2024cll,He:2024iju,Bolton:2024egx,Rosauro-Alcaraz:2024mvx,Eguren:2024oov,Buras:2024ewl,Hati:2024ppg,Allwicher:2024ncl,Becirevic:2024iyi,Altmannshofer:2024kxb,Buras:2024mnq,Hu:2024mgf,Altmannshofer:2025eor,Calibbi:2025rpx,He:2025jfc,Bolton:2025fsq,Chen:2025npb,Aliev:2025hyp,Ding:2025eqq,DiLuzio:2025qkc,Shaw:2025ays}. One approach is to consider that, in addition to the SM neutrinos, the missing energy in the final state could also be carried by other light NP particles, which could naturally explain the Belle~II excess. The scenarios that can resolve the Belle~II excess while satisfying the BaBar and ALEPH upper limits on $\mathcal{B}(B\to K^* E_\mathrm{miss})$ and $\mathcal{B}(B_s\to E_\mathrm{miss})$ are the two-body processes $B \to K \phi / V$~\cite{Altmannshofer:2023hkn} and the three-body processes $B\to K \phi \bar\phi / \psi \bar\psi$~\cite{Fridell:2023ssf,Bolton:2024egx}, where $\phi$, $\psi$, and $V$ are respectively invisible light scalar, spin-1/2, and vector particles.

These light NP particles would also affect other decay processes based on $b \to s E_\mathrm{miss}$ transitions. In this work, we mainly focus on the $\Lambda_b \to \Lambda E_\mathrm{miss}$ decay, which has already attracted some research interest~\cite{Hu:2024mgf,Altmannshofer:2025eor,Das:2025zrn,Lee:2025kvf,Das:2023kch,Aliev:2007rm,Sirvanli:2007yq,Chen:2000mr} and may be measured at the future Tera-Z machines such as FCC-ee~\cite{Amhis:2023mpj}. The Ref.~\cite{Hu:2024mgf} has studied the two-body processes $\Lambda_b \to \Lambda \phi / V$. The present work will carefully investigate the three-body processes $\Lambda_b \to \Lambda \phi \bar\phi / \psi \bar\psi$, presenting the branching ratios and the longitudinal polarization of $\Lambda$ in various NP scenarios, analyzing the correlations between these observables and the branching ratios of $B \to K E_\mathrm{miss}$, as well as examining the differences in contributions from light NP particles to the $\Lambda_b \to \Lambda \phi / V$ and $\Lambda_b \to \Lambda \phi \bar\phi / \psi \bar\psi$ decays.

Our paper is organized as follows. In Sec.~\ref{sec:models}, we present the NP models and the analytical expressions for the contributions of NP to the $\Lambda_b \to \Lambda \phi \bar\phi$ and $\Lambda_b \to \Lambda \psi \bar\psi$ decays. In Sec.~\ref{sec:numerical}, we provide our numerical results and discussions. Our conclusions are made in Sec.~\ref{sec:conclusions}.

\section{Models and observables}
\label{sec:models}

This section presents the models we consider, as well as the branching ratios of $\Lambda_b \to \Lambda E_\mathrm{miss}$ and the longitudinal polarization of $\Lambda$ within these models.

\subsection{The SM}

In the SM, the low-energy effective Hamiltonian governing the $b \to s \nu \bar{\nu}$ transitions is given by 
\begin{align}
\mathcal{H}_\mathrm{eff}^\mathrm{SM} = - \frac{G_F \alpha V_{tb} V_{ts}^*}{\sqrt{2}\pi} C_L^\mathrm{SM} \left(\bar{s} \gamma_\mu P_L b\right) \left( \bar{\nu} \gamma^\mu (1-\gamma_5) \nu \right) + \mathrm{H.c.}.
\label{eq:Heffsm}
\end{align}
Here, $G_F$ is the Fermi constant, $\alpha$ is the fine-structure constant, $V_{tb}$ and $V_{ts}$ are the Cabibbo-Kobayashi-Maskawa (CKM) matrix entries, and the chirality projectors $P_{L,R} = (1 \mp \gamma_5)/2$. Considering the next-to-leading order (NLO) Quantum Chromodynamics (QCD) corrections~\cite{Buchalla:1993bv,Misiak:1999yg,Buchalla:1998ba} and the two-loop electroweak contributions~\cite{Brod:2010hi}, the Wilson coefficient in the SM is $C_L^\mathrm{SM} = -6.32 \pm 0.07$~\cite{Becirevic:2023aov}.

The differential branching ratio for $\Lambda_b \to \Lambda \nu \bar{\nu}$ with the $\Lambda$ baryon in a definite helicity state is given by 
\begin{align}
\frac{d\mathcal{B}\left(\Lambda_b \to \Lambda \nu \bar{\nu} \right)_\mathrm{SM}^\pm}{dq^2} =
& \frac{\tau_{\Lambda_b} G_F^2 \alpha^2 |V_{tb} V_{ts}^*|^2 \sqrt{s_+ s_-}}{512 \pi ^5 m_{\Lambda _b}^3} \left| C_L^\mathrm{SM}\right|^2  \nonumber\\
& \times \Big\{ \Big[ \left(m_{\Lambda_b}+m_{\Lambda}\right)\sqrt{s_-} f_+ \nonumber\\
&\mp \left(m_{\Lambda _b} - m_{\Lambda }\right) \sqrt{s_+} g_+ \Big]^2 \nonumber\\
& + 2 q^2 \left( \sqrt{s_-} f_\perp \mp \sqrt{s_+} g_\perp \right)^2
\Big\}.
\end{align}
Here, the superscript ``$\pm$'' in $d\mathcal{B}\left(\Lambda_b \to \Lambda \nu \bar{\nu} \right)_\mathrm{SM}^\pm$ indicates that the helicities of the final-state $\Lambda$ are $\lambda_\Lambda = \pm 1/2$, respectively. $\tau_{\Lambda_b}$ is the lifetime of $\Lambda_b$ baryon. $s_\pm \equiv \left(m_{\Lambda _b} \pm m_{\Lambda }\right)^2-q^2$ are universal kinematic quantities on the hadronic side.

The differential branching ratio is given by 
\begin{align}
\frac{d\mathcal{B} (\Lambda_b \to \Lambda \nu \bar{\nu})_\mathrm{SM}}{dq^2} 
=&\frac{d\mathcal{B}\left(\Lambda_b \to \Lambda \nu \bar{\nu} \right)_\mathrm{SM}^-}{dq^2} + \frac{d\mathcal{B}\left(\Lambda_b \to \Lambda \nu \bar{\nu} \right)_\mathrm{SM}^+}{dq^2} \nonumber\\
=&\frac{\tau_{\Lambda_b} G_F^2 \left|V_{tb} V_{ts}^*\right|^2 \alpha ^2 \sqrt{s_+ s_-} }{256 \pi ^5 m_{\Lambda _b}^3} \left| C_L^\mathrm{SM}\right|^2  \nonumber\\
&\times \Big\{ s_-\left[ \left(m_{\Lambda _b}+m_{\Lambda }\right)^2 f_+^2 +2 q^2 f_\perp^2\right] \nonumber\\
&+ s_+ \left[ \left(m_{\Lambda _b}-m_{\Lambda }\right)^2 g_+^2 +2 q^2 g_\perp^2 \right] \Big\}.
\end{align}

\subsection{$\Lambda_b \to \Lambda \phi \bar\phi$}

It is assumed that the missing energy associated with the excess observed at Belle~II experiment is carried away by a pair of invisible scalar particles. We write the low-energy effective Hamiltonian suitable for describing the $b \to s \phi \bar\phi$ transitions as 
\begin{align}
\mathcal{H}_\mathrm{eff}^\phi =& \frac{g_{SS}\bar{s}b + g_{PS}\bar{s} \gamma_5 b}{\Lambda} \phi^\dagger \phi \nonumber\\
&+ \frac{g_{VV} \bar{s} \gamma_\mu b + g_{AV} \bar{s} \gamma_\mu \gamma_5 b}{\Lambda^2} i\phi^\dagger \overleftrightarrow{\partial^\mu} \phi + \mathrm{H.c.}.
\label{eq:Heffphi}
\end{align}

For a complex scalar $\phi$, the differential branching ratio for the decay $\Lambda_b \to \Lambda \phi \bar{\phi}$, where the $\Lambda$ baryon is in a definite helicity state, is given by 
\begin{align}
\frac{d\mathcal{B}\left(\Lambda_b \to \Lambda \bar{\phi} \phi \right)^\pm}{dq^2}  =& \frac{n_{\phi}}{8} 
\Bigg(
\frac{12 \Lambda^2}{q^2} \left|l_1 \mp l_2\right|^2 + \beta_2^{\phi 2} \left|l_3 \pm l_4\right|^2 \nonumber\\
&+ 2 \beta_2^{\phi 2} \left|l_5 \pm l_6\right|^2
\Bigg) \label{eq:dbrphi}
\end{align}
with
\begin{align}
n_X \equiv \frac{\tau_{\Lambda_b} \beta^X _2 q^4 \sqrt{s_+ s_-}}{192 \pi ^3 \Lambda ^4 m_{\Lambda _b}^3},
\end{align}
where $\beta^X_2 = \sqrt{1-\frac{4 m_X^2}{q^2}}$. Here, $X$ represents field $\phi$ or $\psi$. The functions $l_{1-6}$ are defined as
\begin{align}
l_1 &=   g_{SS} \kappa_{1+} \kappa_{3-} f_0, &
l_2 &=   g_{PS} \kappa_{1-} \kappa_{3+} g_0, \nonumber\\
l_3 &=   g_{VV} \kappa_{1-} f_+, &
l_4 &=   g_{AV} \kappa_{1+} g_+, \nonumber\\
l_5 &=   g_{VV} \kappa_{2-} f_\perp, &
l_6 &=   g_{AV} \kappa_{2+} g_\perp, \label{eq:gi}
\end{align}
respectively. The kappa variables that contain hadronic kinematic information are respectively given by
\begin{align}
\kappa_{1\pm} = \frac{\left(m_{\Lambda_b} \mp m_\Lambda \right) \sqrt{s_\pm}}{q^2}, 
\kappa_{2\pm} = \sqrt{\frac{s_\pm}{q^2}},
\kappa_{3\pm} = \frac{\sqrt{q^2}}{2(m_b \pm m_s)}.
\end{align}

For a real scalar field, i.e. $\phi = \bar{\phi}$, the contributions from $g_{VV}$ and $g_{AV}$ vanish, and the remaining contributions from $g_{SS}$ and $g_{PS}$ are a factor of 2 larger at the amplitude level. Hence, the replacements $g_{SS} \to 2g_{SS}$ and $g_{PS} \to 2g_{PS}$ must be made in Eq.~\eqref{eq:gi}. The ratio \eqref{eq:dbrphi} should also be multiplied by factor of $1/2$ to account for identical outgoing states.

\subsection{$\Lambda_b \to \Lambda \psi \bar\psi$}

Suppose the missing energy from the excess in the Belle~II experiment is carried by a pair of invisible spin-1/2 particles. The low-energy effective Hamiltonian governing the $b \to s \psi \bar\psi$ transitions is given by
\begin{align}
\mathcal{H}_\mathrm{eff}^\psi = 
&\frac{f_{SS} \bar{s}b + f_{PS} \bar{s} \gamma_5 b}{\Lambda^2} \bar{\psi} \psi
+\frac{f_{SP} \bar{s}b + f_{PP} \bar{s} \gamma_5 b}{\Lambda^2} \bar{\psi} \gamma_5 \psi \nonumber\\
&+ \frac{f_{VV} \bar{s} \gamma_\mu b + f_{AV} \bar{s} \gamma_\mu \gamma_5 b}{\Lambda^2} \bar{\psi} \gamma^\mu \psi \nonumber\\
&+ \frac{f_{VA} \bar{s} \gamma_\mu b + f_{AA} \bar{s} \gamma_\mu \gamma_5 b}{\Lambda^2} \bar{\psi} \gamma^\mu \gamma_5 \psi \nonumber\\
&+ \frac{f_{TT} \bar{s} \sigma_{\mu\nu} b + f_{\tilde{T}T} \bar{s} \sigma_{\mu\nu} \gamma_5 b}{\Lambda^2} \bar{\psi} \sigma^{\mu\nu} \psi + \mathrm{H.c.}.
\label{eq:Heffpsi}
\end{align}

The differential branching ratio for $\Lambda_b \to \Lambda \psi \bar{\psi}$, where $\psi$ is an invisible Dirac fermion and the $\Lambda$ baryon has definite helicity, is given by
\begin{widetext}
\begin{align}
\frac{d\mathcal{B}\left(\Lambda_b \to \Lambda \bar{\psi} \psi \right)^\pm}{dq^2} = n_{\psi} 
\Bigg\{  
&\left| t_1 \mp t_3\right|^2+\left| t_6 \pm t_{12}\right|^2+\left| t_7 \pm t_{13}\right|^2+\left|
t_2 \mp t_4 + t_8 \pm t_{14}\right|^2+\left| t_9 \pm t_{15}\right|^2 \nonumber\\
&+\left| t_{10} \pm t_{16}\right|^2+\left| t_{19} \mp t_{21}\right|^2+\left| t_{20}\mp t_{22}\right|^2+\left| t_{17}\mp t_{23}\right|^2+\left| t_{18} \mp t_{24}\right|^2  \nonumber\\
&+\frac{6 \sqrt{2} \beta _1 }{\beta _3 \beta_4} \Re\left[\left(t_6 \pm t_{12}\right)
\left(t_{17} \mp t_{23}\right)^* + \left(t_7 \pm t_{13}\right) \left(t_{18} \mp t_{24}\right)^*\right]
\Bigg\}.
\end{align}
The functions $t_{1-24}$ are defined as $t_i = f_i  x_i \; (i=1 \cdots 24)$, where
\begin{align}
\vec{f} = \Big\{
&f_{SS} f_0 , f_{SP} f_0 , f_{PS} g_0 , f_{PP} g_0 , f_{VV} f_0 , f_{VV} f_+ , f_{VV} f_\perp , f_{VA} f_0 , \nonumber\\
&f_{VA} f_+ , f_{VA} f_\perp , f_{AV} g_0 , f_{AV} g_+ , f_{AV} g_\perp , f_{AA} g_0 , f_{AA} g_+ , f_{AA} g_\perp , \nonumber\\
&f_{TT} h_+ , f_{TT} h_\perp , f_{TT} \tilde{h}_+ , f_{TT} \tilde{h}_\perp , f_{\tilde{T}T} h_+ , f_{\tilde{T}T} h_\perp , f_{\tilde{T}T} \tilde{h}_+ , f_{\tilde{T}T} \tilde{h}_\perp
\Big\}, \label{eq:vecf}
\end{align}
\begin{align}
\vec{x} = \Big\{
&\sqrt{3} \beta^\psi _2 \kappa_{1+} \kappa_{3-} , \sqrt{3} \kappa_{1+} \kappa_{3-}, \sqrt{3} \beta^\psi _2 \kappa_{1-} \kappa_{3+} , \sqrt{3} \kappa_{1-} \kappa_{3+}, 0 , \frac{\beta _3 \kappa_{1-}}{\sqrt{2}} , \beta _3 \kappa_{2-} , \sqrt{3} \beta _1 \kappa_{1+} , \nonumber\\
&\frac{\beta^\psi _2 \kappa_{1-}}{\sqrt{2}} , \beta^\psi _2 \kappa_{2-} , 0 , \frac{\beta _3 \kappa_{1+}}{\sqrt{2}} , \beta _3 \kappa_{2+} , \sqrt{3} \beta _1 \kappa_{1-} , \frac{\beta^\psi _2 \kappa_{1+}}{\sqrt{2}} , \beta^\psi _2 \kappa_{2+} , \nonumber\\
& \beta _4 \kappa_{2-} , \sqrt{2} \beta _4 \kappa_{1-} , \beta^\psi _2 \kappa_{2+} , \sqrt{2} \beta^\psi _2 \kappa_{1+} , \beta^\psi _2 \kappa_{2-} , \sqrt{2} \beta^\psi _2 \kappa_{1-} , \beta _4 \kappa_{2+} , \sqrt{2} \beta _4 \kappa_{1+} 
\Big\}.
\end{align}
\end{widetext}
The remaining beta variables are
\begin{align}
\beta_1 = \frac{m_\psi}{\sqrt{q^2}},
\beta_3 = \sqrt{1+\frac{2 m_\psi^2}{q^2}}, 
\beta_4 = \sqrt{1+\frac{8 m_\psi^2}{q^2}}, 
\end{align}
respectively.

In the scenario of invisible Majorana fermions, i.e. $\psi = \psi^c$, the result can be obtained from eq.~\eqref{eq:vecf} with the replacements $f_{VV}$, $f_{AV}$, $f_{TT}$, $f_{\tilde{T}T} \to 0$ and $f_{ab} \to 2f_{ab}$ for the remaining couplings. The ratio must also be multiplied by factor of $1/2$ to account for identical outgoing states.

\subsection{The observables}

The effective Hamiltonians \eqref{eq:Heffsm}, \eqref{eq:Heffphi}, and \eqref{eq:Heffpsi} are defined at the typical energy scale $\mu_b \simeq m_b$, where all degrees of freedom above $\mu_b$ are integrated out, including heavy particles in the SM ($W^\pm$, $Z^0$, top quark, and Higgs boson) as well as possible heavy NP particles at scale $\Lambda$. The effects of these heavy degrees of freedom are encoded in the Wilson coefficients: $C_L^\mathrm{SM}$ in \eqref{eq:Heffsm}; the four coefficients $g_{SS}$, $g_{PS}$, $g_{VV}$ and $g_{AV}$ in \eqref{eq:Heffphi}; and the ten coefficients $f_{SS}$, $f_{PS}$, $f_{SP}$, $f_{PP}$, $f_{VV}$, $f_{AV}$, $f_{VA}$, $f_{AA}$, $f_{TT}$ and $f_{\tilde{T}T}$ in \eqref{eq:Heffpsi}. The $f_{0,+,\perp}$, $g_{0,+,\perp}$, $h_{+,\perp}$ and $\tilde{h}_{+,\perp}$ are helicity form factors for the $\Lambda_b \to \Lambda$ transition, and their definitions can be found in Refs.~\cite{Feldmann:2011xf,Detmold:2016pkz}.

When the missing energy in the final state may be carried by pairs of invisible light scalar particles or fermion pairs, the three-body decays produce continuous $q^2$ distributions. Above the production threshold $q^2 = 4 m_X^2$, these distributions differ from the SM prediction. Specifically, the branching ratios satisfy
\begin{align}
\mathcal{B} =
\begin{cases}
\mathcal{B}_\text{SM}, & 0 \leq q^2 < 4 m_X^2 \\
\mathcal{B}_{\text{SM}+X}, & 4 m_X^2 \leq q^2 \leq (m_{\Lambda_b} - m_\Lambda)^2
\end{cases},
\end{align}
where
\begin{align}
\mathcal{B}_{\text{SM}+X} 
&\equiv \mathcal{B}\left(\Lambda_b \to \Lambda E_\text{miss}\right)_{\text{SM}+X} \nonumber\\
&= \mathcal{B}\left(\Lambda_b \to \Lambda \bar{\nu} \nu \right)_\text{SM} + \mathcal{B}\left(\Lambda_b \to \Lambda X \bar{X} \right).
\end{align}

In the $\Lambda_b$ rest frame, the $q^2$-dependent longitudinal polarization asymmetry of $\Lambda$ can be defined by
\begin{align}
P_L^\Lambda(q^2) = \frac{d \mathcal{B}^-/dq^2 - d \mathcal{B}^+/dq^2}{d \mathcal{B}/dq^2},
\end{align}
and the corresponding $q^2$-integrated longitudinal polarization asymmetry is obtained as
\begin{align}
P_L^\Lambda = \frac{\int_{0}^{(m_{\Lambda_b} - m_\Lambda)^2} \left(d \mathcal{B}^-/dq^2 - d \mathcal{B}^+/dq^2 \right) dq^2}{\int_{0}^{(m_{\Lambda_b} - m_\Lambda)^2} \left(d \mathcal{B}/dq^2\right) dq^2}.
\end{align}

\section{Numerical results and discussions}
\label{sec:numerical}

%%%%%%%%%%%%%%%%%%%%%%%%%%%%%%%%%%%%%%%%%%%%
\begin{table}[t]
	\tabcolsep 0.12in
	\renewcommand\arraystretch{1.2}
	\begin{center}
		\caption{\label{tab:input} \small Summary of input parameters used throughout this paper. }
		\vspace{0.18cm}
		\begin{tabular}{ccc} 
			\hline
			Parameter& Value & References
			\\ \hline
			$G_F$ & $1.1663788(6)\times 10^{-5}$ GeV$^{-2}$  &  \cite{ParticleDataGroup:2024cfk}
			\\
			$\alpha$ & $1/128$   &  \cite{ParticleDataGroup:2024cfk}
			\\
			$m_K$ & $493.677(15) \times 10^{-3}$ GeV  &  \cite{ParticleDataGroup:2024cfk}
			\\
			$m_{B^+}$ & $5279.41(7) \times 10^{-3}$ GeV  &  \cite{ParticleDataGroup:2024cfk}
			\\
			$m_{B^0}$ & $5279.72(8) \times 10^{-3}$ GeV  &  \cite{ParticleDataGroup:2024cfk}
			\\
			$m_{\Lambda}$ & $1115.683(6) \times 10^{-3}$ GeV  &  \cite{ParticleDataGroup:2024cfk}
			\\
			$m_{\Lambda_b}$ & $5619.60(17) \times 10^{-3}$ GeV  &  \cite{ParticleDataGroup:2024cfk}
			\\
			$\bar{m}_s(2 \mathrm{GeV})$ & $93.5(8) \times 10^{-3}$ GeV  &  \cite{ParticleDataGroup:2024cfk}
			\\
			$\bar{m}_b(\bar{m}_b)$ & $4.183(7)$ GeV  &  \cite{ParticleDataGroup:2024cfk}
			\\
			$\tau_{B^+}$ & $1.638(4)$ ps  &  \cite{ParticleDataGroup:2024cfk}
			\\
			$\tau_{B^0}$ & $1.517(4)$ ps  &  \cite{ParticleDataGroup:2024cfk}
			\\
			$\tau_{\Lambda_b}$ & $1.471(9)$ ps  &  \cite{ParticleDataGroup:2024cfk}
			\\
			$|V_{tb}|$ & $1.010(27)$   &  \cite{ParticleDataGroup:2024cfk}
			\\
			$|V_{ts}|$ & $41.5(9) \times 10^{-3}$   &  \cite{ParticleDataGroup:2024cfk}
			\\
			$C_L^\mathrm{SM}$ & $-6.32(7)$   &  \cite{Becirevic:2023aov}
			\\
			\hline
			\multicolumn{2}{c}{$B \to K$ form factors} & \cite{Parrott:2022rgu,Bailey:2015dka,Bouchard:2013eph,Gubernari:2023puw}
			\\
			\multicolumn{2}{c}{$\Lambda_b \to \Lambda$ form factors} & \cite{Detmold:2016pkz,Blake:2022vfl}
			\\
			\hline
		\end{tabular}
	\end{center}
\end{table}
%%%%%%%%%%%%%%%%%%%%%%%%%%%%%%%%%%%%%%%%%%%%

%%%%%%%%%%%%%%%%%%%%%%%%%%%%%%%%%
\begin{figure*}[t]
	\centering
	\includegraphics[width=0.45\textwidth]{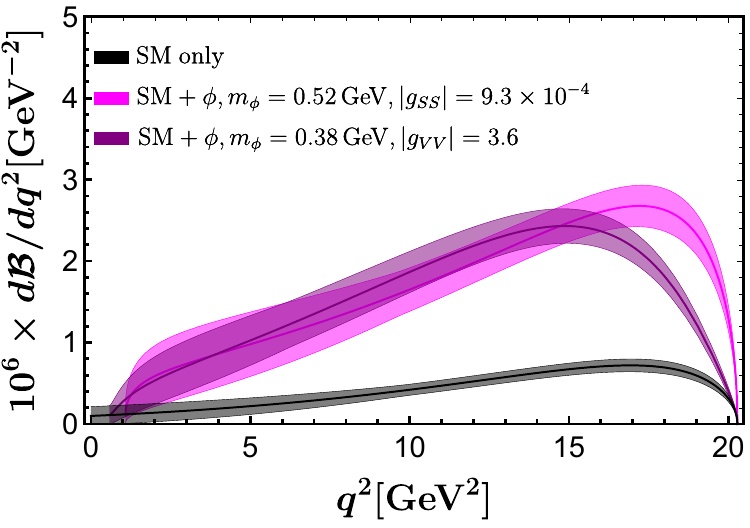}
	\quad
	\includegraphics[width=0.45\textwidth]{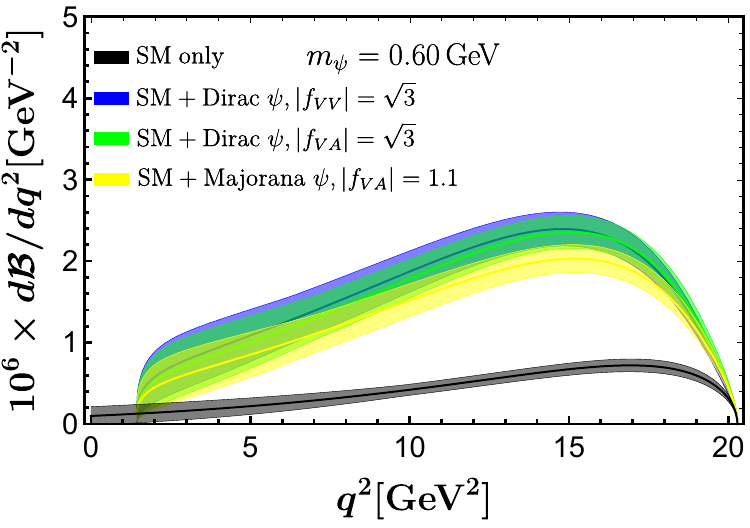}
	\caption{\label{fig:dbr}Predictions for the differential branching ratio of $\Lambda_b \to \Lambda E_\mathrm{miss}$ in the SM and SM+$\phi$ (left panel) as well as the SM and SM+$\psi$ (right panel), using the best-fit mass and couplings from Ref.~\cite{Bolton:2024egx}.}
\end{figure*} 
%%%%%%%%%%%%%%%%%%%%%%%%%%%%%%%%%

%%%%%%%%%%%%%%%%%%%%%%%%%%%%%%%%%
\begin{figure*}[t]
	\centering
	\includegraphics[width=0.45\textwidth]{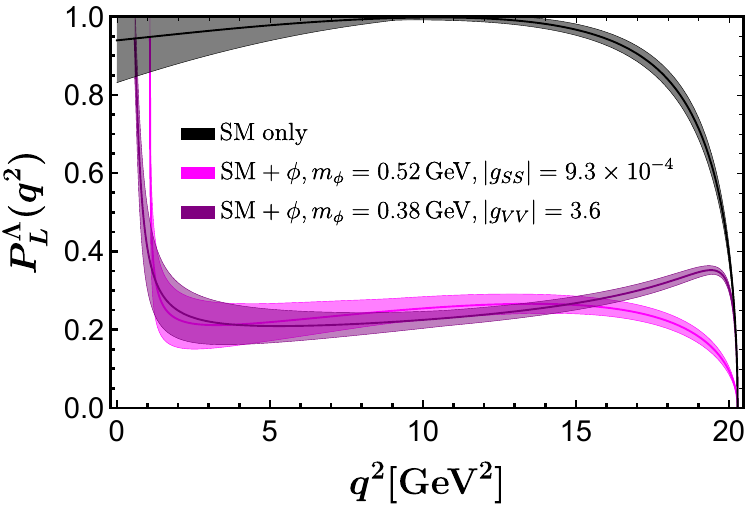}
	\quad
	\includegraphics[width=0.45\textwidth]{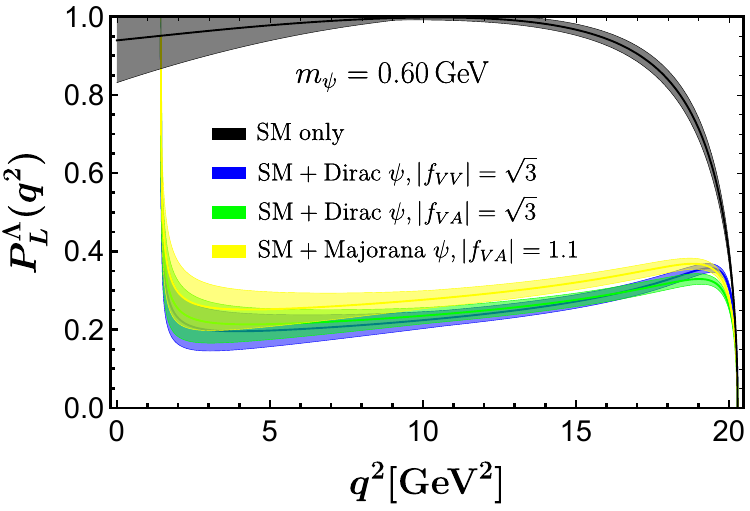}
	\caption{\label{fig:plq2}Predictions for the $q^2$-dependent longitudinal polarization asymmetry of $\Lambda$ in the SM and SM+$\phi$ (left panel) as well as the SM and SM+$\psi$ (right panel), using the best-fit mass and couplings from Ref.~\cite{Bolton:2024egx}.}
\end{figure*} 
%%%%%%%%%%%%%%%%%%%%%%%%%%%%%%%%%

%%%%%%%%%%%%%%%%%%%%%%%%%%%%%%%%%
\begin{figure*}[t]
	\centering
	\includegraphics[width=0.4\textwidth]{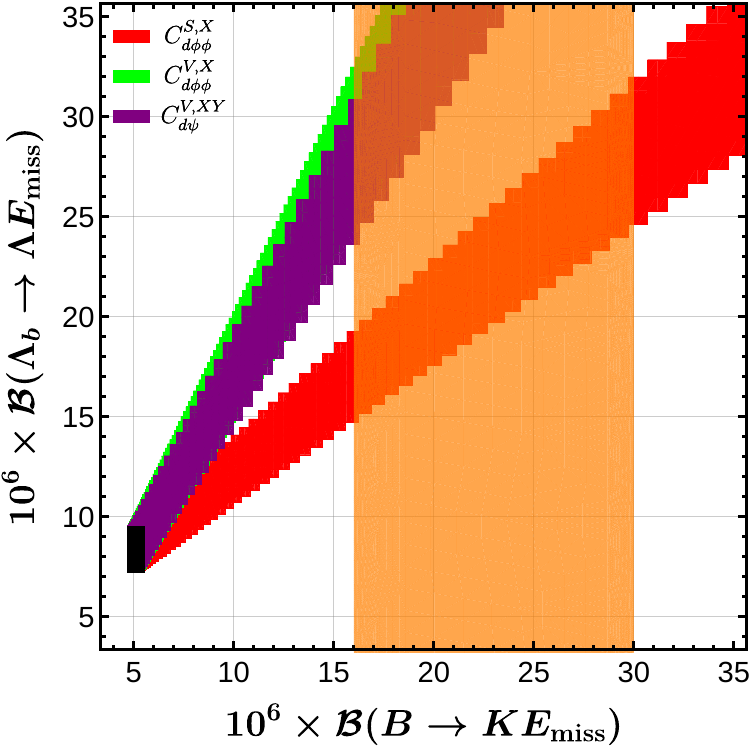}\quad
	\includegraphics[width=0.418\textwidth]{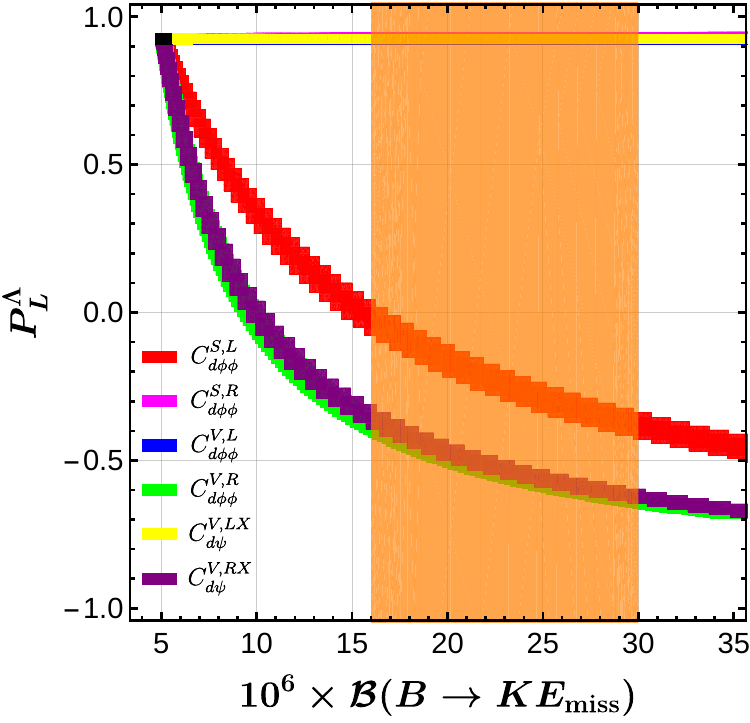}
	\caption{\label{fig:corr}The figure displays the $\mathcal{B}(B \to K E_{\mathrm{miss}})-\mathcal{B}(\Lambda_b \to \Lambda E_{\mathrm{miss}})$ correlation (left panel) and $\mathcal{B}(B \to K E_{\mathrm{miss}})-P^{\Lambda}_{L}$ correlation (right panel) for different NP scenarios. $X,Y \in \{L,R\}$.The SM predictions are represented by black rectangles. The light orange regions indicate the present experimental range~\eqref{eq:BKexp} quoted by Belle-II.}
\end{figure*} 
%%%%%%%%%%%%%%%%%%%%%%%%%%%%%%%%%

Tab.~\ref{tab:input} lists the input parameters used in the numerical analysis, including their associated uncertainties. $\bar{m}_s(2 \mathrm{GeV})$ and $\bar{m}_b(\bar{m}_b)$ are the $\overline{\mathrm{MS}}$ masses renormalized at the scale $2 \mathrm{GeV}$ and $\bar{m}_b$, respectively. In the specific calculation, we use the package RunDec~\cite{Chetyrkin:2000yt,Herren:2017osy} to run the $\bar{m}_s(2 \mathrm{GeV})$ to the typical energy scale $\mu_b$. The overall uncertainties are dominated by those of the transition form factors. Uncertainties from uncorrelated input parameters are combined in quadrature. 

As shown in Fig.~5 of Ref.~\cite{Bolton:2024egx}, in the three-body decay mode, only the $g_{SS}$ and $g_{VV}$ couplings under the scalar pair hypothesis, and the $f_{VV}$ and $f_{VA}$ couplings under the spin-1/2 pair hypothesis, are able to account for the excess observed by the Belle~II experiment. The heavy NP scale is fixed at $\Lambda=10$~TeV throughout this work. 

Fig.~\ref{fig:dbr} shows the contribution of NP effects to the differential branching ratio of $\Lambda_b \to \Lambda E_\mathrm{miss}$, under the assumption that the excess observed by Belle~II arises from a pair of invisible light scalar (left panel) or spin-1/2 (right panel) particles. The black band shows the SM prediction. Assuming the final-state NP particles are invisible light scalar particles, when only the Wilson coefficient $|g_{SS}|=9.3\times 10^{-4}$ is non-zero and the mass of scalar particle takes the corresponding best-fit value $m_\phi=0.52$~GeV, the differential branching ratio of $\Lambda_b \to \Lambda E_\mathrm{miss}$ above the kinematic threshold $q^2 = 4m_\phi^2$ is given by the magenta band; correspondingly, the result for $|g_{VV}|=3.6$ and $m_\phi=0.38$~GeV is represented by the purple band. The prediction for the $g_{SS}$ scenario is larger than that for the $g_{VV}$ scenario at large $q^2$. However, in other regions, large uncertainties from input parameters currently prevent a clear distinction between the two scenarios. Under the invisible light spin-1/2 particles hypothesis, we always take $m_\psi=0.60$~GeV. If $\psi$ is Dirac, the corresponding distribution for the case where only the Wilson coefficient $|f_{VV}|=\sqrt{3}$ is non-zero corresponds to the blue band above the threshold $q^2 = 4m_\psi^2$. Similarly, $|f_{VA}|=\sqrt{3}$ corresponds to the green band. Furthermore, if $\psi$ is Majorana, we take $|f_{VA}|=1.1$, and the result corresponds to the yellow band. The distributions in these three cases cannot be distinguished over the entire kinematic range. Whether they are scalar particles or spin-1/2 particles, their existence leads to a branching ratio for $\Lambda_b \to \Lambda E_\mathrm{miss}$ that is significantly larger than that in the SM only.

Fig.~\ref{fig:plq2} displays the contribution of NP effects to the $q^2$-dependent longitudinal polarization asymmetry of $\Lambda$, assuming that the excess observed by Belle~II originates from a pair of invisible light scalar (left panel) or spin-1/2 (right panel) particles. We study $P_L^\Lambda (q^2)$ within the same NP scenarios as in Fig.~\ref{fig:dbr}, and find that the predicted values of the $P_L^\Lambda (q^2)$ are significantly smaller than those in the SM across all NP scenarios.

Previously, we discussed the analysis in the parity basis. This choice simplifies the analytical expressions for the branching ratios, as the transition form factors for the $\Lambda_b \to \Lambda$ are most naturally defined through hadronic matrix elements with definite parity. However, from a top-down perspective, it is more natural to write the operators in the chiral basis, which couples the light invisible state to $\bar{s} \Gamma P_X b$, where $\Gamma \in \{1,\, \gamma_\mu, \,\sigma_{\mu\nu}\}$ and $X \in \{L,\, R\}$. The parity-basis Wilson coefficients can be converted to those in the chiral basis as
\begin{align}
	\begin{pmatrix}
		g_{SS} \\
		g_{PS} \\
	\end{pmatrix} &= 
	\frac{v}{2\sqrt{2}\Lambda}
	\begin{pmatrix}
		1 & 1 \\
		-1 & 1
	\end{pmatrix}
	\begin{pmatrix}
		C_{d\phi\phi}^{S,L} \\
		C_{d\phi\phi}^{S,R} \\
	\end{pmatrix}, \\
 \begin{pmatrix}
		g_{VV} \\
		g_{AV} \\
	\end{pmatrix} &= 
	\frac{1}{2}
	\begin{pmatrix}
		1 & 1 \\
		-1 & 1
	\end{pmatrix}
	 \begin{pmatrix}
		C_{d\phi\phi}^{V,L} \\
		C_{d\phi\phi}^{V,R} \\
	\end{pmatrix}, \\
\begin{pmatrix}
	f_{VV} \\
	f_{VA} \\
	f_{AV} \\
	f_{AA} \\
\end{pmatrix} &= 
\frac{1}{4}\begin{pmatrix}
	1 & 1 & 1 & 1 \\
	-1 & 1 & -1 & 1 \\
	-1 & -1 & 1 & 1 \\
	1 & -1 & -1 & 1 \\
\end{pmatrix}\begin{pmatrix}
	C_{d\psi}^{V,LL} \\
	C_{d\psi}^{V,LR}  \\
	C_{d\psi}^{V,RL}  \\
	C_{d\psi}^{V,RR}  \\
\end{pmatrix} .
\end{align}

Next, we will discuss the $q^2$-integrated branching ratio $\mathcal{B}(\Lambda_b \to \Lambda E_{\mathrm{miss}})$ and longitudinal polarization asymmetry $P^{\Lambda}_{L}$ in the chiral basis. The correlations between these quantities and the branching ratio of the $B \to K E_{\mathrm{miss}}$ decay are shown in the left panel and right panel of Fig.~\ref{fig:corr}, respectively. It can be seen from the $\mathcal{B}(B \to K E_{\mathrm{miss}})-\mathcal{B}(\Lambda_b \to \Lambda E_{\mathrm{miss}})$ correlation plot that the branching ratio is completely insensitive to the chirality of the operator. Within the measured range of $\mathcal{B}(B \to K E_{\mathrm{miss}})$, scenarios $C_{d\phi\phi}^{V,X}$ and $C_{d\psi}^{V,XY}$ predict similar values for $\mathcal{B}(\Lambda_b \to \Lambda E_{\mathrm{miss}})$, but significantly larger than that predicted by scenario $C_{d\phi\phi}^{S,X}$. This correlation can be used to effectively distinguish between NP scenarios $C_{d\phi\phi}^{V,X}$ (or $C_{d\psi}^{V,XY}$) and $C_{d\phi\phi}^{S,X}$.

The contributions from left-handed and right-handed hadronic current operators exhibit no overlap in the $\mathcal{B}(B \to K E_{\mathrm{miss}})-P^{\Lambda}_{L}$ correlation plot for each NP hypothesis, allowing for clear discrimination. This is similar to the conclusion in the two-body processes $\Lambda_b \to \Lambda \phi / V$~\cite{Hu:2024mgf}. The NP in scenarios $C_{d\phi\phi}^{S,R}$, $C_{d\phi\phi}^{V,L}$, and $C_{d\psi}^{V,LX}$ make an extremely small contribution to $P_L^\Lambda$. In contrast, in scenarios $C_{d\phi\phi}^{S,L}$, $C_{d\phi\phi}^{V,R}$, and $C_{d\psi}^{V,RX}$---which have the opposite hadronic current chirality to the aforementioned scenarios---the NP can rapidly reduce $P_L^\Lambda$; moreover, within the range of the measured value of $\mathcal{B}(B \to K E_{\mathrm{miss}})$, the predicted values of $P_L^\Lambda$ are all negative. This correlation still cannot distinguish between the scenarios $C_{d\phi\phi}^{V,R}$ and $C_{d\psi}^{V,RX}$, but it can more clearly separate scenarios $C_{d\phi\phi}^{V,R}$ (or $C_{d\psi}^{V,RX}$) from $C_{d\phi\phi}^{S,L}$. Furthermore, we find that the $\mathcal{B}(B \to K E_{\mathrm{miss}})-P^{\Lambda}_{L}$ correlation results predicted by scenarios $C_{d\phi\phi}^{V,R}$ and $C_{d\psi}^{V,RX}$ can be clearly separated from the results of the NP scenarios discussed in the two-body processes $\Lambda_b \to \Lambda \phi / V$~\cite{Hu:2024mgf}.

\section{Conclusions}
\label{sec:conclusions}
Recently, the Belle~II measurement of $\mathcal{B}(B \to K E_{\mathrm{miss}})$ exceeds the SM prediction by approximately $2.6\sigma$, sparking considerable interest in NP interpretations. One possible explanation is that the missing energy in the observed excess is carried away by invisible light particles. In this case, the underlying NP would also affect other $b\to s E_{\mathrm{miss}}$ decays.

In this work, we mainly study the influence of invisible light particles on $\Lambda_b \to \Lambda E_{\mathrm{miss}}$, specifically focusing on the decays $\Lambda_b \to \Lambda \phi \bar\phi$ and $\Lambda_b \to \Lambda \psi \bar\psi$. First, in the parity basis, we select the scenarios that can explain the Belle~II excess while satisfying the upper limit constraints from the BaBar and ALEPH experiments on $\mathcal{B}(B\to K^* E_\mathrm{miss})$ and $\mathcal{B}(B_s\to E_\mathrm{miss})$. In these scenarios, we find that the NP can significantly enhance the differential branching ratio $d\mathcal{B}(\Lambda_b \to \Lambda E_{\mathrm{miss}})/dq^2$ and markedly suppress the $q^2$-dependent longitudinal polarization asymmetry of $\Lambda$ over the entire kinematic range above the threshold.

In the chiral basis, we investigate the $\mathcal{B}(B \to K E_{\mathrm{miss}})-\mathcal{B}(\Lambda_b \to \Lambda E_{\mathrm{miss}})$ correlation and $\mathcal{B}(B \to K E_{\mathrm{miss}})-P^{\Lambda}_{L}$ correlation in eight different NP scenarios. The $\mathcal{B}(B \to K E_{\mathrm{miss}})-\mathcal{B}(\Lambda_b \to \Lambda E_{\mathrm{miss}})$ correlation is completely insensitive to the chirality of the effective operators, whereas the $\mathcal{B}(B \to K E_{\mathrm{miss}})-P^{\Lambda}_{L}$ correlation 
shows clear differences between scenarios with different chiralities of the hadronic current. We can use $P_L^\Lambda$ to distinguish the chirality of the hadronic current part in the effective operators, which is similar to the situation in the two-body processes $\Lambda_b \to \Lambda \phi / V$. In addition, we find that both correlations can be used to distinguish scenario $C_{d\phi\phi}^{V,X}$ (or $C_{d\psi}^{V,XY}$) from scenario $C_{d\phi\phi}^{S,X}$. We anticipate more precise measurements of the aforementioned observables, particularly $P^{\Lambda}_{L}$, from experiment such as FCC-ee~\cite{Amhis:2023mpj}. This will help further deepen our understanding of the quark-level $b \to s E_{\mathrm{miss}}$ transitions.

\section*{Acknowledgements}

This work is supported by the National Natural Science Foundation of China under Grant No.~12105002, and the Guangxi Natural Science Foundation under
Grant No.~2023GXNSFBA026270.

%\bibliographystyle{JHEP}
%\bibliography{ref}

\providecommand{\href}[2]{#2}\begingroup\raggedright\endgroup

\end{document}